# Modeling the Role of Context Dependency in the Recognition and Manifestation of Entrepreneurial Opportunity

**Murad A. Mithani**
Rensselaer Polytechnic Institute
Lally School of Management and Technology
110 8th Street, Troy, NY 12180, USA
mitham@rpi.edu

**Tomás Veloz**
University of Chile
Departamento de Ciencias de la Computación
Blanco Encalada 2120 3er Piso Santiago
CHILE
tveloz@dcc.uchile.cl

**Liane Gabora**
University of British Columbia
Okanagan campus, 3333 University Way
Kelowna BC, V1V 1V7
CANADA
liane.gabora@ubc.ca

## Abstract

The paper uses the SCOP theory of concepts to model the role of environmental context on three levels of entrepreneurial opportunity: idea generation, idea development, and entrepreneurial decision. The role of contextual-fit in the generation and development of ideas is modeled as the collapse of their superposition state into one of the potential states that composes this superposition. The projection of this collapsed state on the socio-economic basis results in interference of the developed idea with the perceptions of the supporting community, undergoing an eventual collapse for an entrepreneurial decision that reflects the shared vision of its stakeholders. The developed idea may continue to evolve due to continuous or discontinuous changes in the environment. The model offers unique insights into the effects of external influences on entrepreneurial decisions.

## Introduction

Initiation of an entrepreneurial undertaking begins with the *identification of an opportunity* (Shane and Venkataraman, 2000). A solid knowledge of opportunity identification is critical to understanding entrepreneurial decision-making (Venkataraman, 1997). There are several different kinds and/or components of opportunity identification. Sarasvathy et. al. (2003) describe *opportunity recognition* as finding an existing link in the environment between demand and supply (Hayek, 1945). The second, *opportunity discovery* involves satisfying an existing demand or supply through identification of a new supply or market (Knight, 1921). It is the third, *opportunity creation,* that is most relevant here. It involves the development of a new form of product or service by linking a potential demand to a potential supply (Buchanan and Vanberg, 1991). Entrepreneurial ideas are driven by creative insight, and influenced by beliefs, values, and circumstances, as well as perceptions of the environment (Kirzner, 1997; Langlois, 1984). The entrepreneurial process culminates in an entrepreneurial venture that (hopefully) creates value by offering a new product or process at a price greater than the cost of production (Casson, 1982; Venkataraman, 1997).

The form and identity of opportunity-creating ideas, and the mechanism driving their development, are not well-understood (Miller, 2006). The success of an idea for a business venture is related in an idiosyncratic manner to characteristics of the idea itself as well as to characteristics of the individual who comes up with the idea (Shane and Venkataraman, 2000; Sarasvathy et al, 2003; Schumpeter, 1954; Kirzner, 1997; Baron, 2004). This treatment considers opportunities as ideas that are developed independently by entrepreneurs, which need to be transformed into a successful venture (Kirzner, 1982). Entrepreneurs vary in their ability to extract opportunities (Knight, 1921). This paper proposes environmental context both as a contributor and moderator for the relationship between the individual and the idea. This constitutes a step toward formally modeling the process of opportunity identification as a context-dependent combination.

## Environmental Context and Uncertainty

The environment introduces a substantial amount of uncertainty into the process of opportunity creation. The role of the environment is not completely understood, but it does appear to be of vital importance. Arrow (1974a) writes, "the absence of the market implies that the optimizer faces a world of uncertainty" (pp. 6). While the entrepreneurial actions result in a decrease in the uncertainties affecting the market (Arrow, 1974b, pp. 33), entrepreneurs themselves need to minimize the uncertainty related to their potential actions (Schumpeter, 1934). They attempt to achieve this by maximizing the availability of information relevant to their actions (Kirzner, 1997). This is accomplished in part by considering their ideas in the context of various environmental expectations (Sarasvathy and Venkatarman, 2002, pp. 22). The result is more than a calculation of the optimal alternatives (Baumol, 1993); it is highly dependent on the individuals involved (Kirzner, 1973), and their understanding of the environment (Schumpeter, 1934; Hayek, 1945). Thus, both 'cognitive properties' and 'information corridors' determine the extent of opportunities capitalized upon by an entrepreneur (Shane and Venkataraman, 2000).

During the identification of an opportunity, the situation or context acts as a catalyst in three ways. First, it can help complete the process of forging associations between facts that were already known but the connection between them was not yet recognized, or by transforming a vague association into one that is more concrete (Sarasvathy and Venkataraman, 2000, pp. 8; see also Gabora, 2010). For example, Edison's discovery of the lightbulb came only after many experimenters including Edison himself, had repeatedly failed to make a lightbulb that stayed lit for more than a few minutes. When exposed to the chemical properties of Tungsten, Edison recognized that its conductivity and resistance to oxidation offered the potential to generate longer illumination times. This eventually led to the commercial availability of light bulbs.

Second, the context increases the frequency of ideas that align with it, resulting in a filtering of ideas to those that match individual perception of the environment and appear as workable solutions or developed ideas (Johnson-Laird, 1999). Even though the workable solutions may be impractical or too advanced for their time, the context may give them the credence of viability. As RCA transformed into a highly innovative firm while leading the color television revolution during 1960s, its president, David Sarnoff, shared his vision of a color television on the wall. Because RCA seemed to be in a good position to make it possible, the idea was considered realistic. This led to research in multiple relevant technologies, though it took more than 20 years and the perseverance of Japanese firms to develop the first commercial flat panel display.

Finally, context offers the background that integrates a developed idea with established perspectives, and may generate the social and financial support to increase the idea's probability of success. For example, concern over oil prices and greenhouse gases has led to a higher demand for alternate fuels. This in turn has significantly increased funding by government and private sources for startups focused on such initiatives (Ward, 2004).

Idea generation involves entrepreneurial alertness to possibilities in the environment and their potential to result in useful, aesthetically pleasing, or otherwise valuable new ideas. This has been described cognitively in terms of signal detection (Gaglio and Katz, 2001; Baron, 2004). Idea development occurs as the context triggers conceptual combinations into a workable model (Schumpeter, 1934; Hayek, 1945). The process includes adding, pruning, replacing and extending ideas to fit individual beliefs as well as their perception of the environment. While idea generation is not independent of individual predispositions, the influence is more intrinsic, and incorporates values internalized by the individual over time. Idea development includes the conscious act of mapping them to ordinate values as well as the prevalent perception of the environment (Kirzner, 1997). The influence of external factors on entrepreneurial decision, the behavioral cause of value creation, appears as demand expectations and imitative effects (Venkataraman, 1997; DiMaggio & Powell, 1983). The presence of external support does not guarantee the success of an idea but does increase its feasibility. Cognition factos play a greater role during the early phases, and market information during the latter stages (Shane and Venkataraman, 2000). A summary of the various phases at which context plays a role in entrepreneurial opportunities is provided in Figure 1.

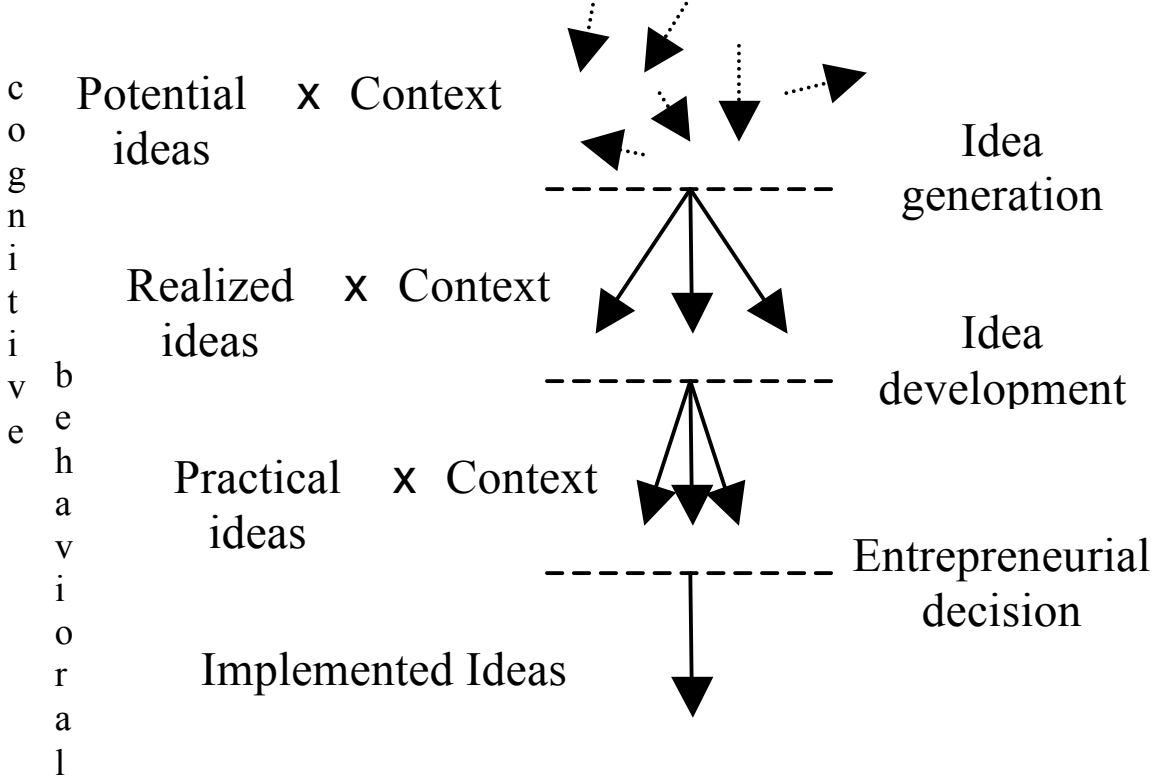

Figure 1. The effect of context at various levels during opportunity identification.

In modeling the environmental context, it is necessary to consider the entrepreneur's perception of the environment, which varies across entrepreneurs due to differences in understanding, beliefs, mood and so forth. Even with respect to individuals experiencing the same external environmental, one may be encouraged by a particular event, and view it as an entrepreneurial opportunity, while others are discouraged by it, viewing it as a problem. It is practical to imagine the perception of the context as lying along the axis between highly positive (very encouraging) and highly negative (very discouraging). By focusing on the individualized perception of the context on entrepreneurial initiatives, we allow for the assignment of subjective values to the context.

## Rationale for the Quantum Approach

Context has long posed a challenge for those attempting to develop formal models in the psychological and social sciences. Linguists and psychologists have struggled to formalize the process by which meaning emerges through interaction of a word, concept, or idea with a context. Because classical models – by which we mean formal models that obey the rules of classical mechanics – do an inadequate job of capturing the shift in meaning that occurs when words or concepts appear in different contexts, or combine in new ways, models derived from the formalisms of quantum mechanics have been tentatively introduced (Aerts and Gabora, 2005a,b; Bruza and Cole, 2005; Bruza, Kitto, Nelson, and McEvoy, 2009; Gabora and Aerts, 2002; Widdows and Peters, 2003; see also the 2009 issue of *Journal of Mathematical Psychology* devoted to this topic). The quantum formalisms are expressly suited to modeling contextuality; indeed they were developed in part to be able to incorporate the change of state that occurs due to the effect of the observer (the context) on the measurement process, and the generation of new states through entanglement. Use of the quantum formalisms to model concept combination leads naturally to use of the quantum formalisms to model creative insights or creative acts, for concept combination is known to lie at the heart of many forms of creativity (Gabora & Aerts, 2009). As an anonymous reviewer of this paper pointed out, in common language when a person is in a process of creation she is said to be "inspired" which comes from "in spiritus" thus appealing to another realm of reality which could be identified with the richer quantum state space not directly accessible by means of measurements. Creativity involves combining concepts or ideas in new ways that often result in the generation of new properties or features. It is this that led us to explore the application of quantum formalisms to modeling the complex interaction between the entrepreneurial ideas and the contexts in which they arise.

## The SCOP Theory of Concepts

A first step toward a formal description of the process by which entrepreneurial opportunity is identified and manifested is to examine the transitions from one cognitive state to another. To algebraically model how context affects these transitions, we make use of the State Context Property (or SCOP) theory of concepts (Aerts, 2009; Aerts and Gabora, 2005a,b; Gabora and Aerts, 2002). The SCOP theory of concepts is particularly well-suited to this task. Like geometrical (Gardenfors, 2000) and dual theory (Wisniewski, 1997) approaches to concepts, the SCOP approach takes into account conceptual structure, but in SCOP the structure of a concept (or combination of them) is influenced in a probabilistic manner by context. Two concepts can be said to be 'connected' to the extent that they are likely to act as contexts for one another and thereby evoke one another in some form.

A SCOP model of a concept consists of a set of *states*, each of which is elicited by a *context*. Each context-specific state is associated with unique sets of *weighted properties*, *exemplar typicalities*, and *transition probability functions*, the last of which give the likelihood that, under a given context, it will undergo a change of state. If a state of a concept is not affected by a particular context, then probability transition function takes the value 1. The state of the concept is said to be an *eigenstate* for this context. Otherwise it is a *potentiality state* for this context, reflecting its susceptibility to change. When one is not thinking of a concept it is in its *ground state,* the concept is always evoked in some context; one never experiences it in its raw or undisturbed ground state. In the ground state typical properties (*e.g*. 'surrounded by water' for the concept ISLAND) have a weight close to 1, while atypical properties (*e.g*. 'square' for ISLAND) have a weight close to 0. From the ground state it is possible to transition to any other state under the influence of some context, and this new state has differently weighted properties. For example, applying the context KITCHEN transforms the ground state of ISLAND to a state where weight or applicability of 'surrounded by water' is (hopefully) low.

The structure of a SCOP is derived by determining the natural relations for sets of states, contexts*,* and properties. If context *e* is more constrained than context *f* (*e.g*. 'in the big box' is more constrained than 'in the box') we say *e* 'is stronger than or equal to' *f*, thereby introducing a partial order relation in the set of contexts *M*. By introducing 'and' and 'or' contexts, *M* obtains the structure of a lattice. By introducing the 'not' context, an orthocomplemented lattice is derived for *M*. The same is done for the set of properties *L*, making it possible to construct a topological representation of a SCOP in a closure space.

To obtain not just qualitative but numerical results, and to model the combination of concepts, the SCOP is embedded in a less abstract, more constrained mathematical structure, complex Hilbert space. A state is described by a unit vector or a density operator, and a context or property by an orthogonal projection. The formalism determines the formulas that describe the transition probabilities between states, and the formulas for calculating the weights of properties, allowing us to predict typicality of exemplars and applicability of properties. The typicalities of exemplars and applicabilities of properties yielded by the calculation matched those obtained in the experiments with human subjects. For example, given a list of questions such as (a) Given context C1, 'The pet is being taught to talk', rate the applicability of the property 'furry', or (b) Given context C2, 'The pet is chewing a bone', rate the typicality of the exemplar 'dog'. Note that we test the prediction that the applicability of *each single property* varies for *each* context, as does the typicality of each exemplar, by an amount that can be calculated from the structure of states and contexts (Aerts and Gabora, 2005b).

### Conceptual Entanglement

The SCOP formalism is able to model combinations of concepts as entangled states by taking the tensor product of the Hilbert spaces of the concept and the context (which may itself be another concept or conglomeration of them). This yields a new state space with states that may exhibit a gain or loss of properties compared to its constituents. The tensor product procedure generates two kinds of vectors: (1) *product states*, for which a change of context can affect one component of the combination and not the other, and (2) *non-product states*, for which a change of context cannot affect one component without affecting the other. Non-product states are the states of entanglement. Another crucial quantum feature taken into account in SCOP is superposition. Both entanglement and superposition, are described in Fock space. (A detailed description of the Fock space model and an explanation of how the SCOP formalism applies to concept formation is given elsewhere (Aerts, 2007, 2009, 2010).

Mathematically, let $I$ be a concept formed by the combination of concepts $C_1$ and $C_2$. Let $v_1$ and $v_2$ be the vectors representing concepts $C_1$ and $C_2$ in the Hilbert space $H$, then the vector $v_I$ representing the concept $I$ will have the form $v_I = ne^{i\theta}(v_1 + v_2) + me^{i\varphi}v_1 \otimes v_2$, where $n, m$ are normalization constants, $\theta, \varphi$ represent the phases of the vectors in the complex Hilbert space. This means $v_I \in H \oplus (H \otimes H)$. This is the representation in Fock space, the direct sum of different tensor products of the Hilbert space describing the states of the individual concepts, which is built in (Aerts, 2007). This Fock space representation models the two fundamental aspects of concept combination, namely (i) the formation of a new concept, due to conceptual emergence, described by the new vector $ne^{i\theta}(v_1 + v_2)$ which is the 'quantum superposition' of the vectors representing the individual concepts $C_1$ and $C_2$, (ii) the product combination of both concepts described by the tensor product vector $me^{i\varphi}(v_1 \otimes v_2)$. It is this 'product combination', which is the classical part, giving rise to the classical logical structure. Although the product part contains a quantum element, since different products states can entangle and form superpositions of products in the tensor product Hilbert space $H \otimes H$. The 'superposition', described in the original Hilbert space $H$ is a pure quantum part, making it possible to describe the emergence of a new concept. When $I$ is formed by the combination of more than two concepts, the Fock space description becomes very complex, since all intermediate cases, where some subset of the concepts give rise to the formation of a new concept by superposition and the remaining subset of concept form into a product vector are all possibilities, which are all faithfully taken into account in this Fock space model.

A test for the presence of superposition and/or entanglement is the violation of Bell-type inequalities (Bell 1964). Bell-type inequalities are tested in general by considering coincidence experiments. In case of the original Bell inequality, one performs four experiments, each with two possible outcomes, which can be performed together (two experiments at the same time) as coincidence experiments. The outcomes are plugged into Bell's inequalities to see whether they are violated. The exists a procedure for demonstrating the violation of Bell's inequalities in cognition (Aerts *et al*. 2000). The procedure entails asking a subject to think about a particular concept and observing how the subject responds to simultaneous manipulations (coincidence experiments) that cause the subject to think of one of two instances of that concept. Plugging the results into Bell's equations demonstrates in theory at least entanglement amongst instances of a concept. Experimental support for the notion that concepts, their combinations, and the relationships between various instances of a concept or a group of related concept can be accurately described as entanglement comes from findings that an 'action at a distance' hypothesis (synchronous activation of a word's associates) supported the results of cued recall experiments better than the 'spreading activation' hypothesis (Nelson, McEvoy, and Pointer 2003).

## A SCOP Model of Entrepreneurial Ideas

Before the idea has been conceived, it is difficult to identify the superpositioned set of ideas within the mind of the entrepreneur, as entrepreneurs themselves are not usually aware of all the ideas and their potential combinations available at any time (Baron, 2004). After the fact, the idea has already come to fruition, and therefore the interaction between it and the context provides little information. The interaction between idea and context causes the amplification of certain elements or versions of the idea, and the disappearance of others (the emergent meaning), thereby creating a partition amongst the set of possible ideas. As a result, some appear as suitable choices mutually exclusive to others that do not materialize in that context. Any moment in a creative entrepreneurial enterprise involves both the current actuality, and a set of potentialities, such that one of these potentialities will become actual in the future. The context may play a key role in this actualization.[1]

### Idea Generation and Development as Conceptual Entanglement and Superposition

In order to use SCOP to understand the first two interactions during opportunity identification, we imagine each idea (potential as well as realized) as a state represented by a vector in the Hilbert space. The set of all ideas that exist at any time appear as a weighted superposition of the states representing each potential idea. With time, changes in individual preferences and values evolve the state of superposition with some ideas replaced by others.

At any time, each state has some transition probability that determines its likelihood to change under a given context and therefore a state of potentiality. The states that remain unchanged under that context are the eigenstates for that context (Gabora, 2007). A potentiality state in one context may be an eigenstate in another. In the absence of a context, two ideas may be entangled, such as figuring out ways to earn a living during idea generation, or, to invest in a technology business versus look for a job when identifying workable solutions during idea development. However, the presence of a context may *collapse* the superposition of states to one (or a few) of them at the cost of the elimination of all others. Leading to an idea composed of only the states for which the context is relevant. This happened with the improvement in Plasma Display Panel (PDP) technology during late 1990s when new ventures entering the industry for the development of various applications of PDP increased while startups in the competing Liquid Crystal Display (LCD) technology decreased.

[1] This was the rationale behind the development of a general evolution a process of *context driven actualization of potential,* or CAP (Gabora and Aerts 2005, 2007), and its corresponding mathematical framework SCOP (Gabora and Aerts 2002a,b, Aerts and Gabora 2005a,b), which we use as the framework for modeling the entrepreneurial process.

## Application of the Formalism

In SCOP, the entrepreneur has a conceptual structure $S$ that incorporates various business concepts. A new idea $I$ is described by a combination of concepts that form a lattice theoretic element of SCOP. This combination incorporates both (i) a *superposition* part that describes an emergent new concept, and (ii) a *product* part that takes into account the classical aspects plus entanglement. Mathematically, $I$ is described by a vector in a Hilbert space formed by the direct sum of tensor products of Hilbert spaces where the forming concepts of $I$ are located. The idea $I$ emerges from the first interaction where the context offers the collapse of the superposed state of potential ideas. In the case of inventor of the flatscreen television David Sarnoff, the concepts of television, wall-hanging frame, aesthetic beauty and convenience may have come together for his wish of a flat TV that could be hung on the wall.

In the presence of a relevant context $C_I$ the idea undergoes a second interaction and appears either as a well-developed opportunity, an improvable idea or a rejectable idea depending on the interaction between the *realized idea* $I$ *and the context* $C_I$. The entrepreneurial evaluation of the convergence process can be modeled by the cost relation $k = \frac{\mu_p(\text{profit }(s_I))}{\mu_c(\text{cost }(s_I))}$ where $\mu$ is the applicability function of the context in which idea $I$ is elicited leading to state $s_I$. If $k > \pi$, we say the process converges, where $\pi$ is the *threshold for good business ideas* based on the perception of the entrepreneur. If $k < \pi$, the entrepreneur must decide if the idea can be improved, by evaluating $k > \xi$, where $\xi$ is a threshold for *bad business ideas.* Of course, it must hold that $0 \leq \xi < \pi \leq 1$.

In SCOP, profit and cost are (like any property in the model) represented by a function depending on the state of the idea (which in turn depends on the context). While we only discuss profit versus cost as the factor involved in deciding the viability of an opportunity, other benefits such as attracting influential clients, opening a new market, creating social benefit, or other costs such as damaging the environment, losing influence on the market, and so forth, could also be part of the trade-off in a general model of entrepreneurship. The process of developing the idea during the second interaction can be decomposed into two parts:

1) Given the idea, find new (or modify the known) contexts in which the idea is realizable in a potentially cheaper (cost optimizing) way.
2) Given a context, find new (or modify the known) ideas that are potentially more profitable.

The domain of the profit-applicability function emerging during idea development must be extended to the next stage (the entrepreneurial decision stage) for its completion. If the cross-fertilized process described by 1 and 2 converges to a state/context such that the state is an eigenstate of the context (i.e. there is no change of state driven by the context), and there is sufficient profit margin relative to the cost, the new idea is recognized as a good fit to the environment. As discussed later, in most cases the fit only emerges after the idea is negotiated with the stakeholders and shapes into a shared vision of the entities involved. If the cost relation is not encouraging at the end of any stage, or does not lead to a state for the given context in a reasonable time, or a deadlock is encountered (i.e. there is no profitable avenue for the venture given the available resources), it may result in an abandonment of the idea. It is possible that a change in external conditions may render an abandoned idea feasible in the future. Abandonment will therefore only be an outcome of the evaluation of all potential scenarios relevant to developing the idea into a workable solution. This implicitly assumes that entrepreneurial ideas are never truly abandoned.

The process of convergence can be summarized as:

a) If the identified context appears to offer a cost relation $k > \pi$ then the process converges. The entrepreneur considers it a practical idea and takes it to the next stage.
b) If the cost relation produces $k < \pi$ where $k > \xi$, the entrepreneur intuits that the business idea has not reached its best form. This can undergo a further development as in step 2.
c) If the cost relation results in $k < \pi$ where $k < \xi$, the entrepreneur may decide to abandon the idea.

Let us make this more concrete by returning to the flatscreen television example. As RCA researchers experimented with the available technologies for a wall hanging television during the 1960s, they found LCD and Light Emitting Diode (LED) to be promising candidates. The identification of two out of the many potential display technologies represents the collapse of an idea into workable solutions.

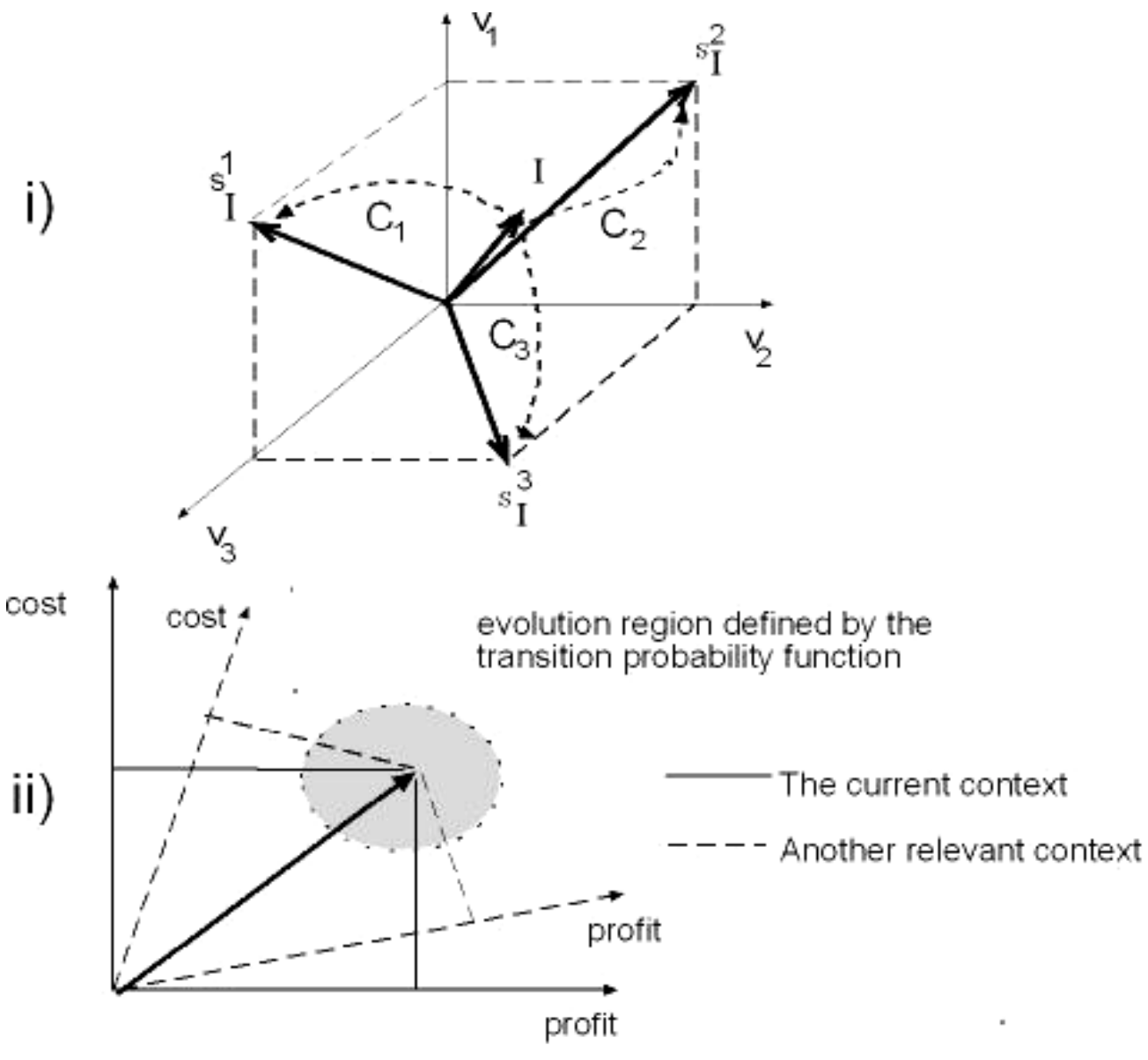

Figure 2. (i) Illustrative example of the collapse process of idea I to states $s_I^1$, $s_I^2$, and $s_I^3$ due to contexts $C_1$, $C_2$, and $C_3$, represented by the conceptual bases $v_1$, $v_2$, and $v_3$. (ii) Idea evolution occurs through evaluation of the profit-cost function in potentially relevant contexts using the probability transition function.

## The Interaction between Venture Choices and Context

The contextual-fit of developed ideas to the community of stakeholders requires extending SCOP to include the projection of a state along the dimensions representing the resources necessary for the venture. Each social and economic resource appears as a separate dimension in this basis (such as finance, personnel, industry contacts, etc.). The collapsed state from the previous phase appears as one of the potential states in addition to the states representing the perception of various stakeholders that include family, friends, financial institutions, and so forth. The process of communication and sharing leads to a superposition of states arising from their interference. States aligned with environmental demands along the various axes have a greater projection along those dimensions and a higher chance of implementation. In other words, a timely idea meeting the expectations of the various support groups will show significant projection along the basis, while a highly unusual idea will appear orthogonal. As the entrepreneur makes a decision, a collapse of the superpositioned state is forced upon this socioeconomic basis that represents the combined vision of the entrepreneur(s) and their supporting community. In most cases, it also incorporates the vision of those who did not or could not participate into the venture through their criticism and feedback. In certain cases, the idea may remain unchanged due to the conviction, persuasive power and/or the financial capacity of the entrepreneur, appearing as an eigenstate in the socioeconomic bases. Returning once again to the television example, of the three display technologies, RCA focused on extensive development and subsequent use of LCDs as commercial products. While most of its leading researchers disagreed, RCA sales divisions focused on Point of Sale (POS) displays as its primary market. The researchers behind this technology disagreed with this vision and went out to develop their own start-ups, initiating the large scale commercialization of the LCD technology in the US (Castellano, 2005).

The entrepreneurial decision may undergo further evolution before resulting in a venture. Such an evolution may be the result of changes in the context that may affect its feasibility. These contextual changes may be continuous or discontinuous. As an example of continuous change, the gradual change of attitude toward start-up proposals developing Internet-based technologies during the 1990s resulted in an unprecedented increase in the magnitude of support during that period. Discontinuous change, on the other hand, appears as a change in the basis relevant to the idea. For example, there is currently an increased focus on investments in alternate fuel technologies for automobiles, but those working in improving the steam and electric engines during the early part of the last century went out of business with the improvements in the efficiency of the gasoline engine (e.g. Miner et al, 2001). An independent change in the basis may turn a failed business model into a lucrative opportunity.

## Significance and Future Work

Efforts are underway to develop a new framework for understanding what is arguably the most uniquely human characteristic: the capacity to invent, and the propensity to put our own spin on others' inventions such that novelty accumulates. This paper presents a preliminary application of a generalization of the quantum formalisms to model how contextual factors enter and affect entrepreneurial ventures. In the future we plan to tie the model to existing theories of creativity from applied psychology, such as Osborn's (1953) Creative Problem Solving method, which involves exploring the vision, formulating challenges, exploring ideas, formulating a solution, exploring acceptance, and formulating a plan. We will also apply the model to Cooperrider and Srivastra's (1987) Appreciative Inquiry technique, which involves identifying and expanding on what is working well as opposed to focusing on what is not working. Both techniques have been the subject of substantial investigation since they were originally introduced. Finally, efforts will be made to expand the model to encompass ideas arising in organizational contexts, and to match output of the model with entrepreneurial data.

## Acknowledgements

This work is funded by grants to Gabora from the Social Sciences and Humanities Research Council of Canada (SSHRC) and the GOA program of the Free University of Brussels. We thank Diederik Aerts for comments on the manuscript.